\documentclass[%
 reprint,
 superscriptaddress,
 amsmath,amssymb,
 aps,
]{revtex4-2}

\usepackage{graphicx}
\usepackage{dcolumn}
\usepackage{bm}

\begin{document}

\preprint{APS/123-QED}

\title{Turbulence-induced clustering in compressible active fluids}

\author{Vasco M. Worlitzer}%
 \email{vasco.worlitzer@ptb.de}
\affiliation{%
Department of Mathematical Modelling and Data Analysis, Physikalisch-Technische Bundesanstalt Braunschweig und Berlin, Abbestrasse 2-12, D-10587 Berlin, Germany
}%
\author{Gil Ariel}%
\affiliation{%
Department of Mathematics, Bar-Ilan University, 52900 Ramat Gan, Israel
}%

\author{Avraham Be'er}%
\affiliation{%
Zuckerberg Institute for Water Research, The Jacob Blaustein Institutes for Desert Research, Ben-Gurion University of the Negev, Sede Boqer Campus, 84990 Midreshet Ben-Gurion, Israel
}%
\affiliation{Department of Physics, Ben-Gurion University of the Negev, 84105 Beer Sheva, Israel}

\author{Holger Stark}%
\affiliation{%
Institute of Theoretical Physics, Technische Universität Berlin, Hardenbergstrasse 36, D-10623 Berlin, Germany
}%

\author{Markus Bär}%
\affiliation{%
Department of Mathematical Modelling and Data Analysis, Physikalisch-Technische Bundesanstalt Braunschweig und Berlin, Abbestrasse 2-12, D-10587 Berlin, Germany
}%

\author{Sebastian Heidenreich}%
\affiliation{%
Department of Mathematical Modelling and Data Analysis, Physikalisch-Technische Bundesanstalt Braunschweig und Berlin, Abbestrasse 2-12, D-10587 Berlin, Germany
}%

\date{\today}

\begin {abstract}
 We study a novel phase of active polar fluids, which is characterized by the continuous creation and destruction of dense clusters due to self-sustained turbulence. This state arises due to the interplay of the self-advection of the aligned swimmers and their defect topology. The typical cluster size is determined by the characteristic vortex size. Our results are obtained by investigating a continuum model of compressible polar active fluids, which incorporates typical experimental observations in bacterial suspensions by assuming a non-monotone dependence of speed on density.
\end{abstract}

\maketitle


Active matter has become a central topic of contemporary physics with applications ranging from collective motion of animals and cells \cite{Vicsek2012} to new forms of soft matter \cite{Marchetti2013} and the dynamics of cells and tissues \cite{Prost2015}. Among active systems, bacterial colonies provide an intriguing and experimentally well-controllable class of systems. Many aspects of their behavior can be understood by analogies to fundamental non-equilibrium statistical physics \cite{Beer2019}. Several bacterial species provide showcases for collective motion and ordered patterns, that can be described by simple models of self-propelled particles dominated by alignment \cite{Chate2020} or repulsive interactions and particle shape \cite{Bar2020}. Suspensions of swimming bacteria may be governed, in addition, by longer range hydrodynamic interactions, which can lead to a new dynamic state referred to as meso-scale turbulence. This dynamical state is characterized by the presence of self-sustained vortices and jets and is reminiscent of turbulence \cite{Dombrowski2004, Cisneros2007, Ishikawa2011, Liu2012, Steager2008, Grossmann2014}. In contrast to inertial turbulence there exists a characteristics vortex size \cite{Dombrowski2004, Zhang2009, Ishikawa2011, Wensink2012a}, which can be explained by the competition between alignment and hydrodynamic interactions \cite{Wensink2012a, Dunkel2013b, Heidenreich2016, Reinken2018a}. However, models of meso-scale turbulence commonly assume incompressibility \cite{Wensink2012a, Dunkel2013b, Heidenreich2016, Reinken2018a}, and are therefore not able to describe clustering phenomena.

A common description of compressible active fluids is based on self-propulsion and purely repulsive interactions showing intriguing non-equilibrium phenomena like motility-induced phase separation (MIPS) \cite{Fily2012, Bialke2013, Cates2015}. For this phenomenon it is crucial that the self-propulsion speed of the particle slows down with increasing density \cite{Bialke2013}. In this context, experiments of bacterial suspensions exhibit a puzzling situation: On the one hand, the speed of bacteria increases with density for low and intermediate densities \cite{Ariel2018} and on the other hand, clusters or density variations are commonly present and often emerge from dilute initial conditions \cite{Beer2020}. This behavior has not been explained so far.

\begin{figure}
	\centering
	\includegraphics[width=\linewidth, height=\textheight,keepaspectratio]{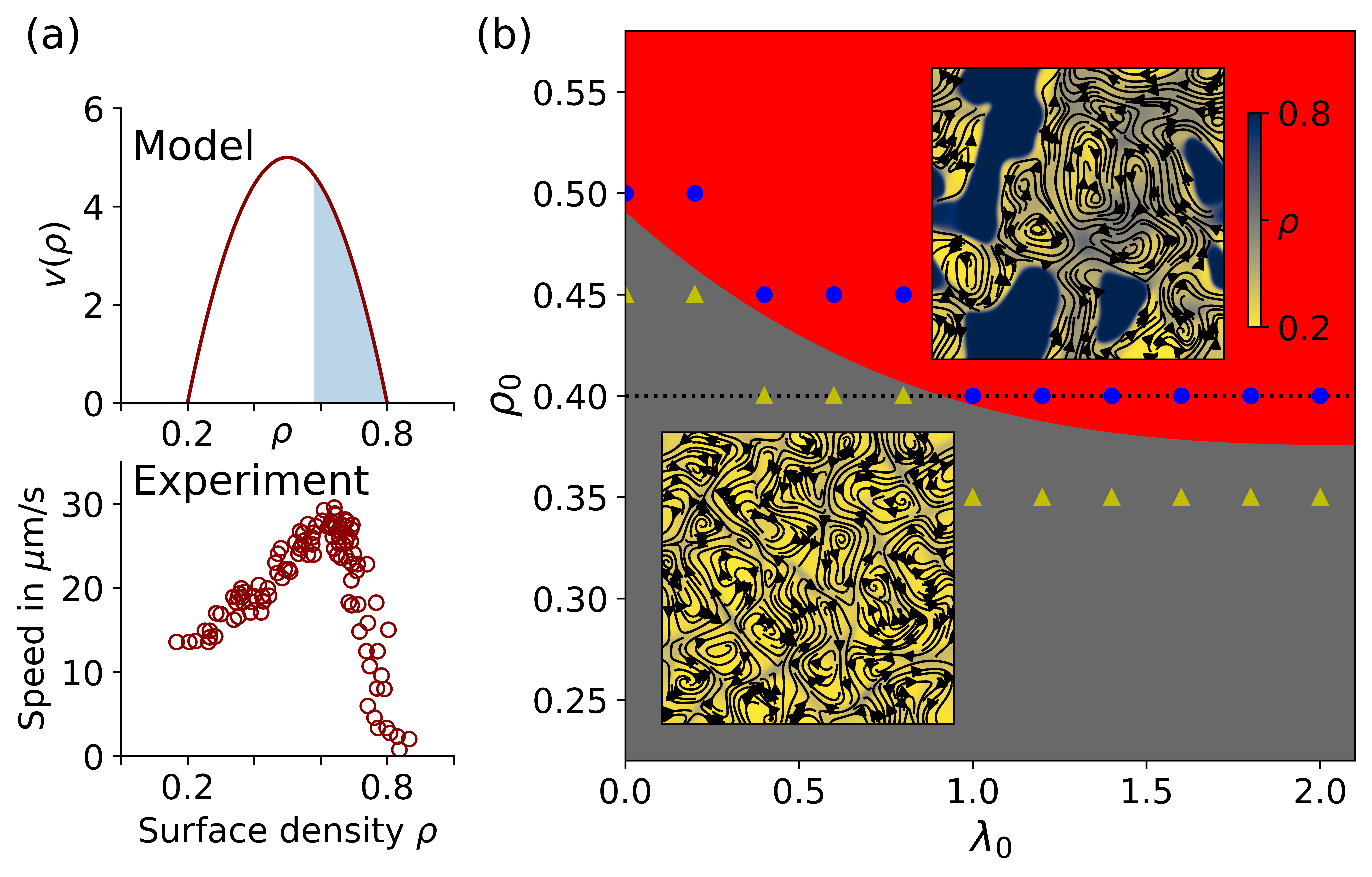}
	\caption{(a) Density-dependent speed $v(\rho)$ in our model (see eq.\ \eqref{eq:v_rho}) and obtained from experiments (reproduced from \cite{Ariel2018}). Linearly unstable region due to MIPS is colored in blue \cite{supp}. (b) Phase portrait for $\rho_0$ and $\lambda_0$ exhibiting meso-scale turbulence with homogeneous density (MST, grey) and turbulence-induced clustering (TIC, red). Symbols represent simulation data, from which the phase boundary is estimated (yellow triangles = MST, blue circles = TIC). Insets: Typical snapshots of the respective states. Streamlines calculated from $\mathbf{v}$, while the background color is obtained from $\rho$ (see colorbar). For better visualization no streamlines are plotted for areas with $\rho=\rho_{max}$, i.e.\ inside dense clusters.}
	\label{fig:phase_portrait}
\end{figure}

Based on our previous investigations \cite{Worlitzer2021}, where a continuum model was proposed that combines meso-scale turbulence with MIPS, we extend here this concept and incorporate typical experimental observations in bacterial suspensions by assuming a non-monotone dependence of speed on density, see fig.\ \ref{fig:phase_portrait}(a). Through a numerical study of the proposed continuum model, we discover the novel dynamical state of turbulence-induced clustering, which is characterized by the continued formation, reshaping and fracture of dense clusters. We show that the dynamic arises due to self-advection of the aligned swimmers and point out that the nucleation of clusters is organized by the defect topology in the polar order field of the swimmers. In general, topological defects often dominate the dynamics of active biological matter: they govern cell death and extrusion in epithelial tissues \cite{Saw2017}, control the dynamics in neural cell cultures \cite{Kawaguchi2017} and promote new layer formation in dense bacterial colonies \cite{Copenhagen2021}. Moreover, we show that the average cluster size depends linearly on the characteristic vortex size. Overall, our results point out an alternative way how clusters can emerge in systems far from equilibrium.

The proposed model for compressible active fluids consists of coupled equations for the density $\rho$ and the polarization density $\mathbf{p}$, quantifying polar order, see \cite{Worlitzer2021}
\begin{subequations}\label{eq:Dynamics}
	\begin{align}\label{eq:rho_dynamics}
	\partial_t \rho &= -\nabla\cdot[v(\rho)\mathbf{p}]+D\Delta\rho,\\
	\label{eq:p_dynamics}
	\partial_t \mathbf{p} +\lambda_0 (\mathbf{p}\cdot\nabla)\mathbf{p}
	=&-\frac{1}{2}\nabla [v(\rho)\rho] - [A(\rho)+C|\mathbf{p}|^2]\mathbf{p} \notag \\  & +\Gamma_0\Delta\mathbf{p}-\Gamma_2\Delta^2\mathbf{p}.
	\end{align}
\end{subequations}
The evolution of the density is governed by the continuity equation eq.\ \eqref{eq:rho_dynamics}, where the current density consists of a diffusive part and a term stemming from the directed motion of the particles. Eq.\ \eqref{eq:p_dynamics} is reminiscent of the equation to model meso-scale turbulence \cite{Heidenreich2016, Reinken2018a}. The terms arise due to self-propulsion, alignment and  hydrodynamic interactions. The pressure-like term involving the gradient of $\rho$ couples the polarization to the evolution of the density, and the velocity field $\mathbf{v}(\mathbf{x},t)$ is obtained from $\mathbf{v} = v(\rho)\mathbf{p}/\rho$. We model the density-dependent speed $v(\rho)$ with the quadratic function
\begin{equation}\label{eq:v_rho}
    v(\rho) = -c\left[\rho-(\rho_{max}+\rho_{min})/2\right]^2 + v_0,
\end{equation}
where $c=4v_0/(\rho_{max}-\rho_{min})^2$. This form is motivated by experimental studies of bacterial suspensions \cite{Sokolov2007, Ariel2018}, see fig.\ \ref{fig:phase_portrait}(a). We remark that the behaviour of bacterial suspensions at the onset of meso-scale turbulence, i.e.\ in the limit $\rho\to\rho_{min}$, is poorly understood. Hence, we do not expect to accurately capture the dynamics with our model in this limit.

Furthermore, density $\rho$ and polarization $\mathbf{p}$ couple to each other via the polar alignment strength $A(\rho)$ in eq.\ \eqref{eq:p_dynamics}. Inspired by Landau theory, one often assumes $A(\rho) = A_0(\rho_c-\rho)$ to model the polar-isotropic phase transition. For simplicity, we assume that the onset of collective motion coincides with the polar-isotropic phase transition, i.e.\ $\rho_c=\rho_{min}$ and hence $A(\rho)<0$.
 
In our numerical simulations, we choose $\rho_{min}=0.2$ and $\rho_{max}=0.8$ in line with experimental results \cite{Ariel2018}. Moreover, we fix $A_0=0.3$, $C=0.5$, $D=5$ and $v_0=5$. These parameters do not affect the dynamics significantly. We vary the mean density $\rho_0$, and use $\Gamma_0$ and $\Gamma_2$ to control the characteristic vortex size. Finally, the strength of self-advection of the polarization is tuned by $\lambda_0$ (for details on the numerics see \cite{supp}).


We initiate simulations with a homogeneous density profile and choose $\Gamma_0=-1$ and $\Gamma_2=1$, such that the system is in a meso-scale turbulent state \cite{Worlitzer2021, supp}. This state is characterized by vortices with a characteristic length scale as well as the presence of dynamical jets \cite{Oza2016a, Wensink2012a, Dunkel2013b}. By varying the mean density $\rho_0$ and the strength of self-advection $\lambda_0$, we sketch a phase portrait, see fig.\ \ref{fig:phase_portrait}(b). Note that the mean density $\rho_0$ is varied outside the spinodal regime of MIPS, i.e.\ between $\rho_{min}$ and $\rho_s\approx 0.58$, see fig.\ \ref{fig:phase_portrait}(a) and \cite{supp}.

Meso-scale turbulence (MST) or vortex lattices accompanied by an almost constant density $\rho\approx\rho_0$ are observed for low values of $\rho_0$ and $\lambda_0$. For high values of $\rho_0$ and $\lambda_0$, we discover a novel phase, which we refer to as turbulence-induced clustering (TIC). In this phase dense clusters with $\rho=\rho_{max}$, hence $\mathbf{v}=0$, emerge throughout the simulation domain, while dilute areas are still characterized by vortices. The dense clusters are not fixed in space but rather get reshaped and stretched. Furthermore, clusters commonly fracture and smaller parts of the fracture might vanish. We emphasize that the emergence, reshaping and fracture of clusters continues indefinitely. That is, while a cluster might fracture in one part of the simulation domain, a new cluster might emerge at another place, see the supplemental movie \cite{supp}.

To quantify the dynamical evolution of dense clusters, we use Minkowski functionals. They provide a mathematical framework to completely characterize the morphology of patterns \cite{Mecke2000, Sofonea1999, supp}. In two dimensions, the Minkowski functionals coincide with the area $A$, perimeter $U$ and Euler number $\chi$. The Euler number can be calculated by counting the number of connected finite domains, in our case the dense clusters, and subtracting the number of holes. Hence, the Euler number is a topological quantity related to the number of clusters. Minkowski functionals provide a useful tool to study the highly irregular shapes of dense clusters in our system, while also allowing for comparison with the more commonly studied quantities such as the number of clusters and the characteristic domain size derived from the density correlation function \cite{supp, Mecke2000}.

\begin{figure}
	\centering
	\includegraphics[width=\linewidth, height=\textheight,keepaspectratio]{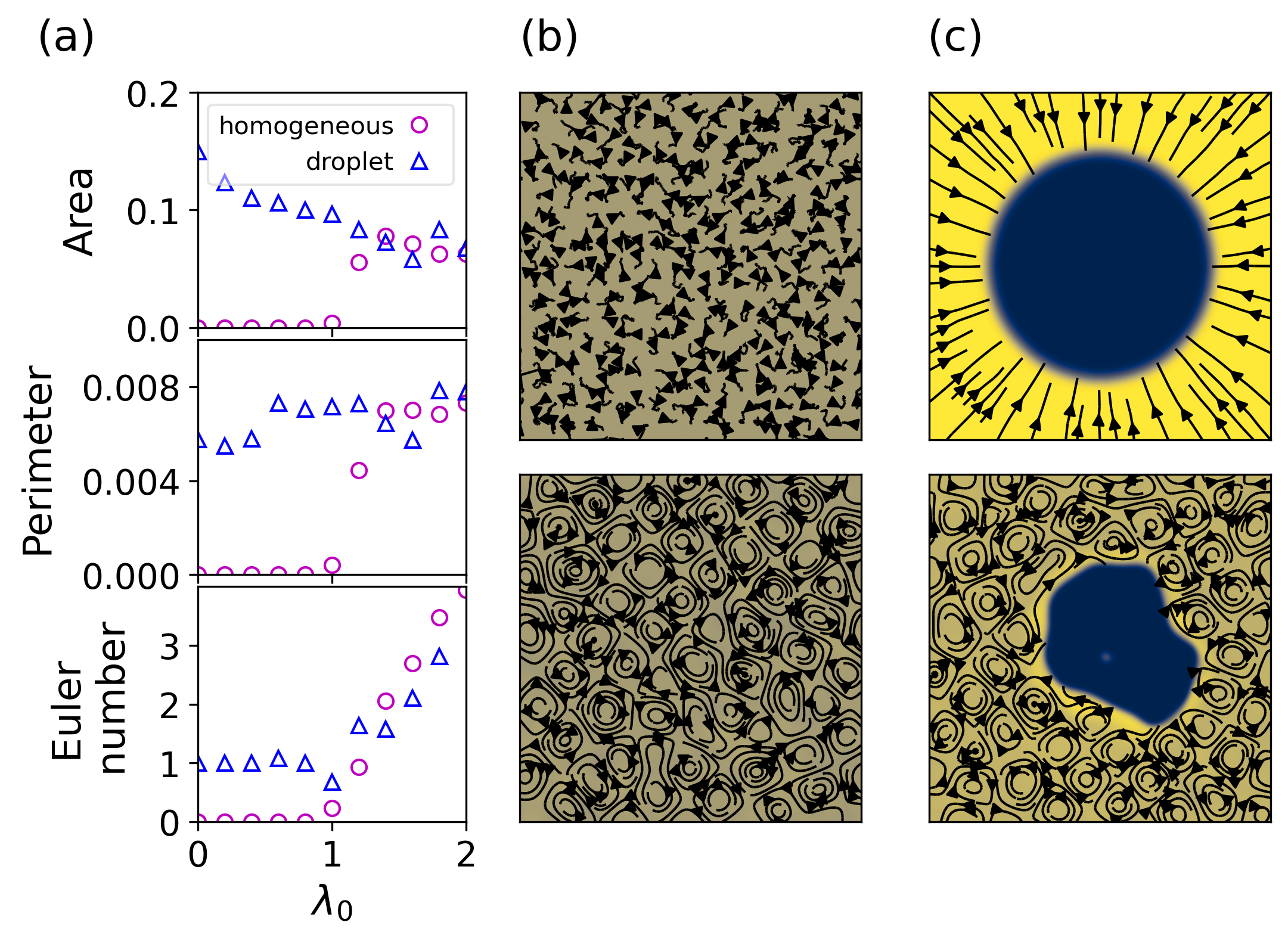}
	\caption{(a) Time-averaged Minkowski functionals when varying $\lambda_0$ along the dotted line in fig.\ \ref{fig:phase_portrait}(b), i.e.\ for $\rho_0=0.4$. Circles represent runs started from a homogeneous density profile, while triangles represent runs started from an initial dense droplet. Surface area and perimeter are rescaled by the domain size to yield fractions. (b), (c) Snapshots of the initial condition (upper) and at $t=150$ (end of the simulation, lower) for a homogeneous density profile (b) and an initial droplet (c) in the MST phase, i.e.\ for $\rho_0=0.4$ and $\lambda_0=0$. Respective snapshots of the TIC phase are provided in \cite{supp}.}
	\label{fig:Minkowski_dynamics}
\end{figure}

Analyzing the temporal evolution of the Minkowski functionals in the TIC phase reveals that $A$, $U$ and $\chi$ fluctuate around a non-zero mean value \cite{supp} due to continuous creation and destruction of clusters, which indicates a statistical steady state. We rule out finite-size effects by varying the linear size $L$ of the system \cite{supp}. To further test the universality of our results, we initiate simulations with a single dense circular domain and vary the self-advection strength $\lambda_0$ along the dotted line in fig.\ \ref{fig:phase_portrait}(b). For low values of $\lambda_0$ (i.e.\ in the MST phase) the mean values of the Minkowski functionals are well separated (fig.\ \ref{fig:Minkowski_dynamics}(a)): An initial circular cluster preserves its Euler number 1 and maintains positive area and perimeter, while the Minkowski functionals are zero for the homogeneous density profile. Hence, the time evolution in the MST phase depends heavily on the initial conditions, see fig.\ \ref{fig:Minkowski_dynamics}(b) and (c). For high values of $\lambda_0$ (i.e.\ in the TIC phase) differences in the time-averaged Minkowski functionals cannot be attributed to the initial conditions. After a transient, the dynamics and statistical properties of the system for different initial conditions are indistinguishable, see fig.\ \ref{fig:Minkowski_dynamics}(a) and \cite{supp}. From this observation we can draw two conclusions about the TIC phase: 1) The novel statistical steady state is independent of the initial conditions and 2) the completely phase separated state is unstable. The second conclusion in turn shows that another process than MIPS must be responsible for the TIC phase, as MIPS predicts a complete phase separation on long time scales. Our results indicate that self-advection through the non-linear term $\lambda_0(\mathbf{p}\cdot\nabla)\mathbf{p}$ of the polarization triggers the nucleation of clusters as well as the fracture of already existing domains.

\begin{figure}
	\centering
	\includegraphics[width=\linewidth, height=\textheight,keepaspectratio]{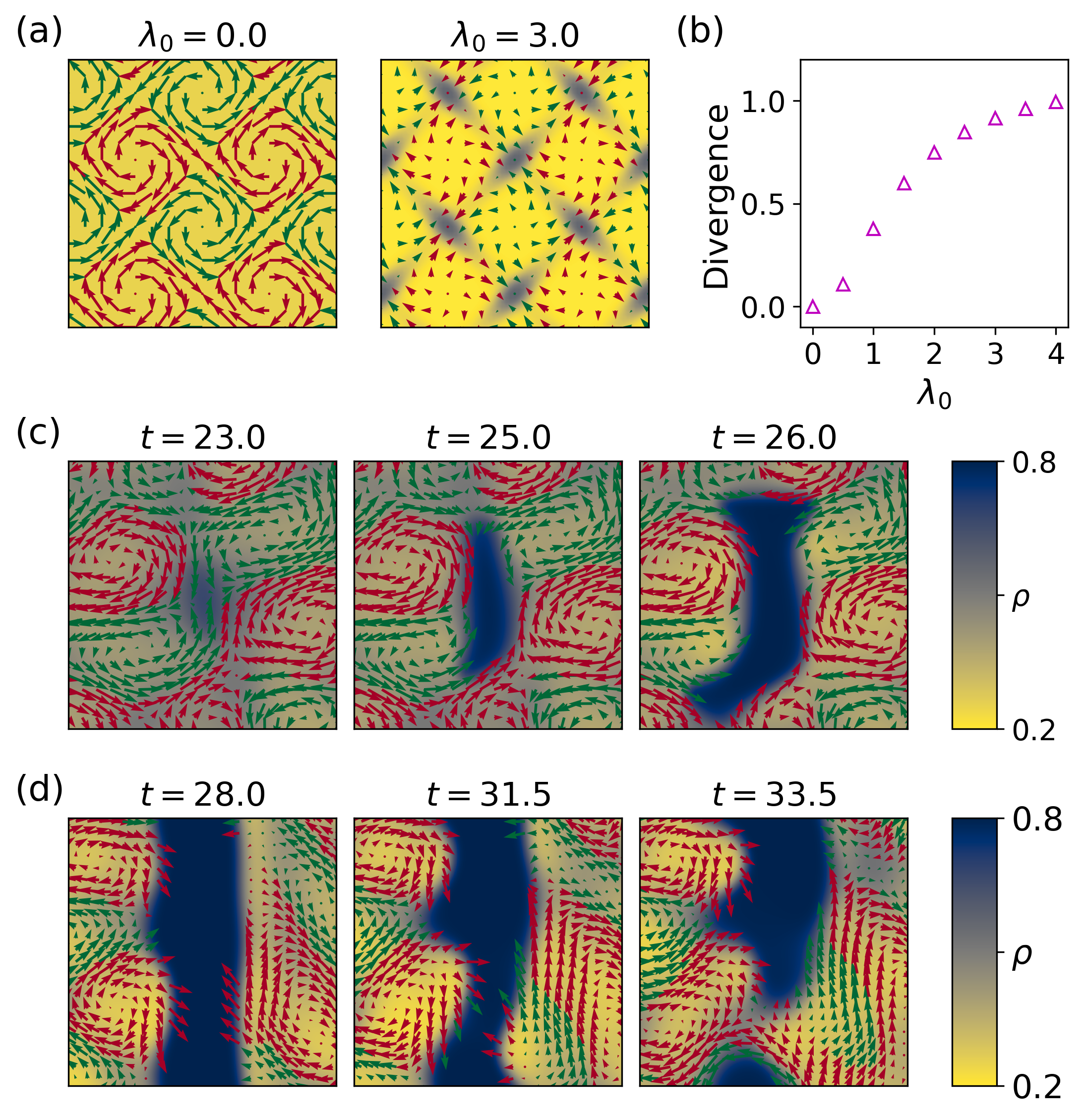}
	\caption{(a) Snapshots at $t=20$ of runs started from a vortex lattice with $\rho_0=0.25$, $\Gamma_0=-2$ for $\lambda_0=0$ and $\lambda_0=3$. (b) Maximal absolute divergence, i.e.\ $\max_{[0,\delta t]}1/L^2\int (\nabla\cdot\mathbf{p})^2 d\mathbf{x}$ for the initial stage of the vortex lattice when varying $\lambda_0$. Parameters as in (a). (c) Snapshots of a cluster growing from random initial conditions with $\rho_0=0.45$, $\lambda_0=0.5$. (d) Evolution of a flat interface between a dense cluster and a vortex array in the dilute region with $\rho_0=0.45$ and $\lambda_0=3$. In (a), (c) and (d) arrows represent the $\mathbf{p}$ field, with arrow color indicating clockwise (red) and counterclockwise (green) rotating vortices.}
	\label{fig:snapshots}
\end{figure}

The nucleation of clusters due to meso-scale turbulence can be understood by the compressibility of the system and the local defect topology. We illustrate this by initiating the system with a vortex lattice at constant density $\rho=\rho_0$. In this situation, the evolution of the density eq.\ \eqref{eq:rho_dynamics} reduces to $\partial_t \rho = -v(\rho_0)\nabla\cdot\mathbf{p}$. Hence, density changes are only possible if sinks ($\nabla\cdot\mathbf{p}<0$) or sources ($\nabla\cdot\mathbf{p}>0$) of the polarization are present. We remark that this argument is only valid at the early stages, when density gradients are negligible. We detect sources and sinks of the polarization by computing the integral $1/L^2 \int (\nabla\cdot\mathbf{p})^2 \;d\mathbf{x}$. Fig.\ \ref{fig:snapshots}(b) shows that the maximum of this integral increases with the strength of self-advection $\lambda_0$, leading to density gradients within the system. 

Analyzing the snapshots in fig.\ \ref{fig:snapshots}(a) shows that the density changes are organized by the defect topology of the polarization. In a vortex lattice, a topological defect with charge 1 is located at the vortex center and the point in between four adjacent vortices is a topological defect with charge -1. While for $\lambda_0=0$ the vortex lattice is only perturbed marginally, high values of $\lambda_0$ result in the appearance of sources of the polarization at the vortices. Clearly, combining a source with a vortex results in an outward spiral. This is in line with simulation results from \cite{Elgeti2011}. Therein, elongated pushers (like \textit{Bacillus subtilis}) show a propensity to form outward spirals, when increasing self-advection. Density shifted away from the vortices accumulates in between four adjacent vortices (fig.\ \ref{fig:snapshots}(a)) due to mass conservation. This is reminiscent of the results of \cite{Torney2007}, where swimmers forced by a velocity field in the form of a vortex lattice leave the vortices and accumulate in between them. Hence, there is a density shift from topological defects with charge 1 to those with topological charge -1, which represent possible nucleation sites for dense clusters. Indeed, in simulations started from random initial conditions, clusters preferentially nucleate in between four adjacent vortices, see fig.\ \ref{fig:snapshots}(c). Furthermore, initiating the system with a flat interface between a dense cluster and a vortex array in the dilute region, shows that the surrounding vortices modulate and eventually break-up the interface, see fig.\ \ref{fig:snapshots}(d).

Finally, we address the mechanism which determines the average cluster size in the TIC phase. Trivially, the cluster size depends on the mean density $\rho_0$ \cite{supp}. More interestingly, for fixed $\rho_0$, the cluster size changes with the characteristic vortex size. The characteristic vortex size can be estimated from the fastest growing mode $k_c$ obtained from the stability analysis as $\Lambda=2\pi/k_c=2\pi\sqrt{2\Gamma_2/-\Gamma_0}$ \cite{supp}.

\begin{figure}
	\centering
	\includegraphics[width=\linewidth, height=\textheight,keepaspectratio]{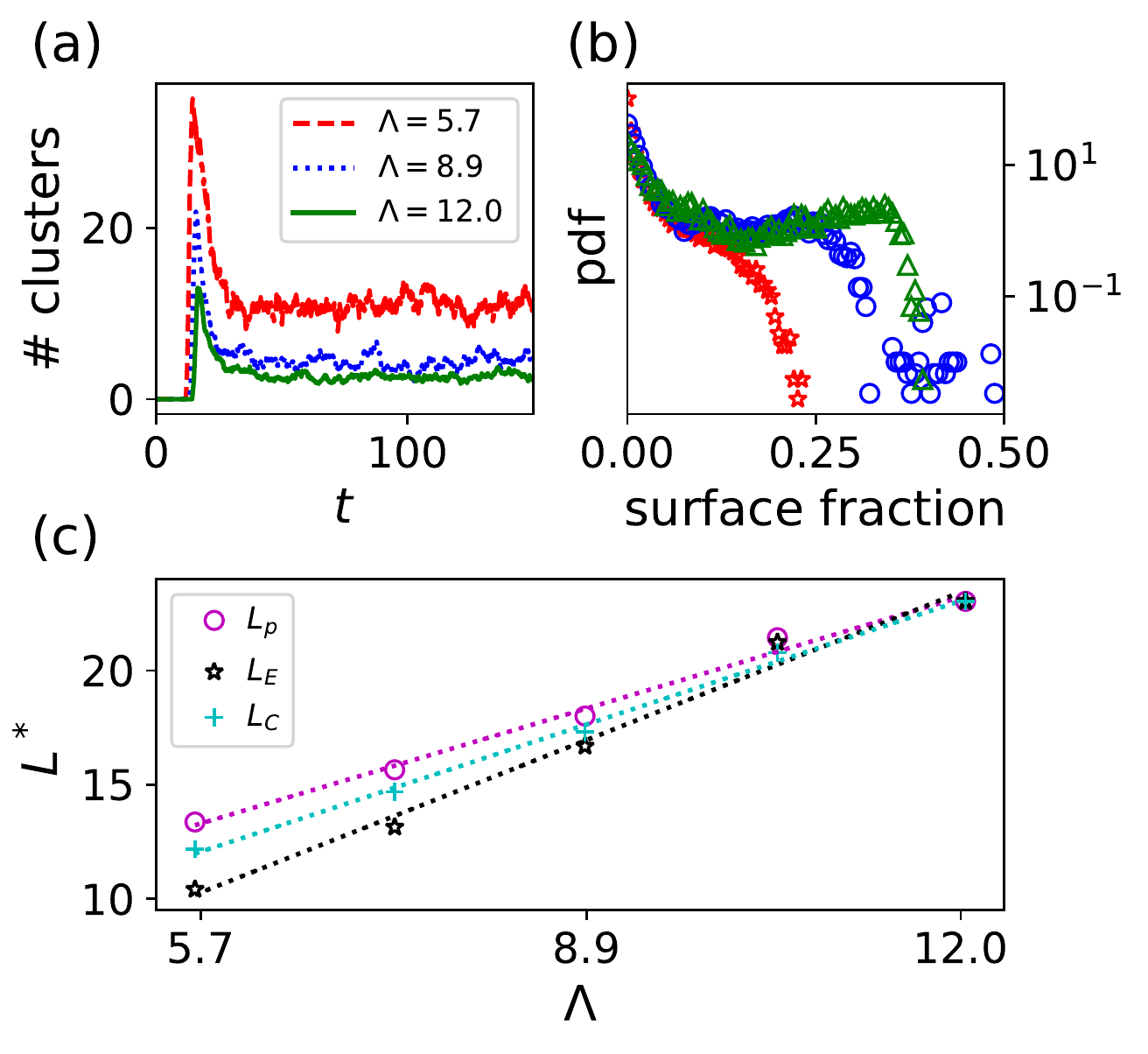}
	\caption{(a) Number of clusters, (b) cluster size distribution and (c) characteristic domain size $L^*$ for different characteristic vortex sizes $\Lambda=2\pi\sqrt{2\Gamma_2/-\Gamma_0}$. Details for the choice of $\Gamma_0,\Gamma_2$ can be found in the supplemental material \cite{supp}. Data is obtained for $\rho_0=0.5$ by averaging over 10 runs ((a),(b)) and 5 runs (c) with identical parameters but different random initial conditions. Colors in (b) correspond to those in (a). In (c) the characteristic domain size obtained from the perimeter ($L_p$, magenta circles), the Euler number ($L_E$, black stars) and the correlation function ($L_c$, cyan crosses) is shown. As a visual guide linear fits for all three cases are provided as dashed lines. $L_p$ and $L_E$ are rescaled with the maximum of $L_c$.}
	\label{fig:coarsening_length}
\end{figure}

Increasing $\Lambda$ results in a decrease in the number of clusters (fig.\ \ref{fig:coarsening_length}(a)), but a higher probability to find larger clusters, which can be deduced from the cluster size distribution plotted in fig.\ \ref{fig:coarsening_length}(b). Moreover, we compute a characteristic domain size $L^*$ either from the density correlation function ($L_c$), the perimeter ($L_p$) or the Euler number ($L_E$) \cite{supp}. All three quantities increase linearly with $\Lambda$, see fig.\ \ref{fig:coarsening_length}(c). This shows that the mean value, around which the Minkowski functionals fluctuate, is determined by the characteristic vortex size. A possible explanation is provided by our previous observation that clusters preferentially nucleate in between adjacent vortices. Increasing the vortex size leaves more space in between vortices (implies larger distance between topological defects with charge +1), while the number of possible nucleation sites reduces, resulting in fewer but larger clusters. Hence, we conjecture that the novel TIC phase can be understood as fluctuations around a vortex lattice with dense cluster situated in between vortices.

We presented an extension of a continuum model that combines MIPS with meso-scale turbulence by including a realistic self-propulsion speed dependence on density. By numerical investigations, we showed that meso-scale turbulence induces dynamical clustering if incompressibility is not enforced. Independent of the system size and initial conditions, a novel dynamical state emerges, which is characterized by the continuous nucleation, deformation and destruction of clusters. These antagonistic dynamics arise due a common cause: Hydrodynamic interactions lead to self-advection of the polarization governing the dynamics. This points to the importance of hydrodynamic interactions, which are usually excluded in the literature \cite{Jayaram2020, Grossmann2020}, when studying clustering phenomena of elongated self-propelled particles.

Furthermore, the TIC phase extends to densities $\rho_0>\rho_{s}$, i.e.\ to the regime where spinodal decomposition through MIPS is expected: Clusters form initially through a process reminiscent of spinodal decomposition, while on long time scales we observe continued emergence, reshaping and fracture of clusters \cite{supp}. Hence, turbulence-induced clustering is a generic phenomena for incompressible polar active fluids.

Our results bear some reminiscence to passive systems quenched into the spinodal regime in the presence of turbulence. Therein, thermodynamic forces drive coarsening, while existing domains are broken up due to the turbulent motion. The competition between these effects leads to a coarsening arrest at a certain length scale \cite{Lacasta1995, Berthier2001, Berti2005, Perlekar2014}. However, there are two notable novel aspects of our observations: Firstly, nucleation of clusters due to inertial turbulence is not reported for passive systems \cite{Lacasta1995, Berthier2001, Berti2005, Perlekar2014}, where clusters emerge spontaneously as the system is quenched into the spinodal regime. Our simulations are performed outside the spinodal regime. Consequently, clustering is initiated and governed by the turbulent fluid motion in the active system. Secondly, the average cluster size in the active system is controlled by the characteristic vortex size, rather than by opposing forces. Hence, turbulence-induced clustering illustrates a novel route to pattern formation which is unique to active systems. In the context of biological systems, it points to a possible functional benefit of meso-scale turbulence, e.g.\ as a driver for aggregation of bacteria at low and intermediate densities.

This work was supported by the Deutsche Forschungsgemeinschaft (DFG) through grants HE 5995/3-1 (SH, VMW and AB), BA 1222/7-1 (MB and GA) and SFB 910 (projects B4 (HS) and B5 (MB)). GA and AB are thankful for partial support from the Israel Science Foundation grant 373/16.

\bibliography{all}



\end{document}